\def\ba{\begin{array}}
	\def\ea{\end{array}}
\def\be{\begin{equation}}
\def\ee{\end{equation}}
\def\bg{\begin{aligned}}
	\def\eg{\end{aligned}}
\documentclass[pra,aps,showpacs,twocolumn,twoside,superscriptaddress]{revtex4}

\usepackage{amsmath,amsfonts,amssymb,caption,color,epsfig,graphics,graphicx,hyperref,latexsym,mathrsfs,revsymb,theorem,url,verbatim,epstopdf}

\usepackage{amssymb}
\usepackage{color}
\usepackage{amsmath,epic,curves,amscd}
\usepackage[english]{babel}
\usepackage{graphicx}
\usepackage{comment}
\usepackage{appendix}
\usepackage{mathdots}

\newtheorem{definition}{Definition}
\newtheorem{proposition}[definition]{Proposition}
\newtheorem{lemma}[definition]{Lemma}

\newtheorem{theorem}[definition]{Theorem}
\newtheorem{corollary}[definition]{Corollary}
\newtheorem{conjecture}[definition]{Conjecture}

\newtheorem{remark}[definition]{Remark}
\newtheorem{example}[definition]{Example}
\newtheorem{question}[definition]{Question}

\def\squareforqed{\hbox{\rlap{$\sqcap$}$\sqcup$}}
\def\qed{\ifmmode\squareforqed\else{\unskip\nobreak\hfil
		\penalty50\hskip1em\null\nobreak\hfil\squareforqed
		\parfillskip=0pt\finalhyphendemerits=0\endgraf}\fi}
\def\endenv{\ifmmode\;\else{\unskip\nobreak\hfil
		\penalty50\hskip1em\null\nobreak\hfil\;
		\parfillskip=0pt\finalhyphendemerits=0\endgraf}\fi}
\newenvironment{proof}{\noindent \textbf{{Proof.~} }}{\qed}
\def\Dbar{\leavevmode\lower.6ex\hbox to 0pt
	{\hskip-.23ex\accent"16\hss}D}
\makeatletter
\def\url@leostyle{%
	\@ifundefined{selectfont}{\def\UrlFont{\sf}}{\def\UrlFont{\small\ttfamily}}}
\makeatother
\def\bcj{\begin{conjecture}}
	\def\ecj{\end{conjecture}}
\def\bcr{\begin{corollary}}
	\def\ecr{\end{corollary}}
\def\bd{\begin{definition}}
	\def\ed{\end{definition}}
\def\bea{\begin{eqnarray}}
\def\eea{\end{eqnarray}}
\def\bem{\begin{enumerate}}
	\def\eem{\end{enumerate}}
\def\bex{\begin{example}}
	\def\eex{\end{example}}
\def\bim{\begin{itemize}}
	\def\eim{\end{itemize}}
\def\bl{\begin{lemma}}
	\def\el{\end{lemma}}
\def\bma{\begin{bmatrix}}
	\def\ema{\end{bmatrix}}
\def\bpf{\begin{proof}}
	\def\epf{\end{proof}}
\def\bpp{\begin{proposition}}
	\def\epp{\end{proposition}}
\def\bqu{\begin{question}}
	\def\equ{\end{question}}
\def\br{\begin{remark}}
	\def\er{\end{remark}}
\def\bt{\begin{theorem}}
	\def\et{\end{theorem}}

\def\bea{\begin{eqnarray}}
\def\eea{\end{eqnarray}}
\def\beq{\begin{equation}}
\def\eeq{\end{equation}}
\def\bal{\begin{aligned}}
	\def\eal{\end{aligned}}
\def\bma{\begin{bmatrix}}
	\def\ema{\end{bmatrix}}

\def\btb{\begin{tabular}}
	\def\etb{\end{tabular}}

\newcommand{\nc}{\newcommand}

\def\a{\alpha}
\def\b{\beta}
\def\g{\gamma}
\def\d{\delta}

\def\l{\lambda}

\def\s{\sigma}

\nc{\bbA}{\mathbb{A}} \nc{\bbB}{\mathbb{B}} \nc{\bbC}{\mathbb{C}}
\nc{\bbD}{\mathbb{D}} \nc{\bbE}{\mathbb{E}} \nc{\bbF}{\mathbb{F}}
\nc{\bbG}{\mathbb{G}} \nc{\bbH}{\mathbb{H}} \nc{\bbI}{\mathbb{I}}
\nc{\bbJ}{\mathbb{J}} \nc{\bbK}{\mathbb{K}} \nc{\bbL}{\mathbb{L}}
\nc{\bbM}{\mathbb{M}} \nc{\bbN}{\mathbb{N}} \nc{\bbO}{\mathbb{O}}
\nc{\bbP}{\mathbb{P}} \nc{\bbQ}{\mathbb{Q}} \nc{\bbR}{\mathbb{R}}
\nc{\bbS}{\mathbb{S}} \nc{\bbT}{\mathbb{T}} \nc{\bbU}{\mathbb{U}}
\nc{\bbV}{\mathbb{V}} \nc{\bbW}{\mathbb{W}} \nc{\bbX}{\mathbb{X}}
\nc{\bbZ}{\mathbb{Z}}

\nc{\bA}{{\bf A}} \nc{\bB}{{\bf B}} \nc{\bC}{{\bf C}}
\nc{\bD}{{\bf D}} \nc{\bE}{{\bf E}} \nc{\bF}{{\bf F}}
\nc{\bG}{{\bf G}} \nc{\bH}{{\bf H}} \nc{\bI}{{\bf I}}
\nc{\bJ}{{\bf J}} \nc{\bK}{{\bf K}} \nc{\bL}{{\bf L}}
\nc{\bM}{{\bf M}} \nc{\bN}{{\bf N}} \nc{\bO}{{\bf O}}
\nc{\bP}{{\bf P}} \nc{\bQ}{{\bf Q}} \nc{\bR}{{\bf R}}
\nc{\bS}{{\bf S}} \nc{\bT}{{\bf T}} \nc{\bU}{{\bf U}}
\nc{\bV}{{\bf V}} \nc{\bW}{{\bf W}} \nc{\bX}{{\bf X}}
\nc{\bZ}{{\bf Z}}

\nc{\cA}{{\cal A}} \nc{\cB}{{\cal B}} \nc{\cC}{{\cal C}}
\nc{\cD}{{\cal D}} \nc{\cE}{{\cal E}} \nc{\cF}{{\cal F}}
\nc{\cG}{{\cal G}} \nc{\cH}{{\cal H}} \nc{\cI}{{\cal I}}
\nc{\cJ}{{\cal J}} \nc{\cK}{{\cal K}} \nc{\cL}{{\cal L}}
\nc{\cM}{{\cal M}} \nc{\cN}{{\cal N}} \nc{\cO}{{\cal O}}
\nc{\cP}{{\cal P}} \nc{\cQ}{{\cal Q}} \nc{\cR}{{\cal R}}
\nc{\cS}{{\cal S}} \nc{\cT}{{\cal T}} \nc{\cU}{{\cal U}}
\nc{\cV}{{\cal V}} \nc{\cW}{{\cal W}} \nc{\cX}{{\cal X}}
\nc{\cY}{{\cal Y}}
\nc{\cZ}{{\cal Z}}

\nc{\hA}{{\hat{A}}} \nc{\hB}{{\hat{B}}} \nc{\hC}{{\hat{C}}}
\nc{\hD}{{\hat{D}}} \nc{\hE}{{\hat{E}}} \nc{\hF}{{\hat{F}}}
\nc{\hG}{{\hat{G}}} \nc{\hH}{{\hat{H}}} \nc{\hI}{{\hat{I}}}
\nc{\hJ}{{\hat{J}}} \nc{\hK}{{\hat{K}}} \nc{\hL}{{\hat{L}}}
\nc{\hM}{{\hat{M}}} \nc{\hN}{{\hat{N}}} \nc{\hO}{{\hat{O}}}
\nc{\hP}{{\hat{P}}} \nc{\hR}{{\hat{R}}} \nc{\hS}{{\hat{S}}}
\nc{\hT}{{\hat{T}}} \nc{\hU}{{\hat{U}}} \nc{\hV}{{\hat{V}}}
\nc{\hW}{{\hat{W}}} \nc{\hX}{{\hat{X}}} \nc{\hZ}{{\hat{Z}}}

\nc{\hn}{{\hat{n}}}

\def\max{\mathop{\rm max}}

\def\tr{\mathop{\rm Tr}}

\def\ox{\otimes}

\newcommand{\bra}[1]{\langle#1|}
\newcommand{\ket}[1]{|#1\rangle}
\newcommand{\proj}[1]{| #1\rangle\!\langle #1 |}

\begin{document}

\title{Trade-off Relation among Genuine Three-qubit Nonlocalities in Four-qubit Systems}



\author{Li-Jun Zhao}\email[]{lijunzhao@buaa.edu.cn}
\affiliation{School of Mathematics and Systems Science, Beihang University, Beijing 100191, China}
\author{Lin Chen}\email[]{linchen@buaa.edu.cn (corresponding author)}
\affiliation{School of Mathematics and Systems Science, Beihang University, Beijing 100191, China}
\affiliation{International Research Institute for Multidisciplinary Science, Beihang University, Beijing 100191, China}
\author{Yu-Min Guo}\email[]{guo_ym@cnu.edu.cn (corresponding author)}
\affiliation{School of Mathematical Sciences, Capital Normal University, Beijing 100048, China}
\author{Kai Wang}
\affiliation{School of Mathematics and Systems Science, Beihang University, Beijing 100191, China}
\author{Yi Shen}
\affiliation{School of Mathematics and Systems Science, Beihang University, Beijing 100191, China}
\author{Shao-Ming Fei}\email[]{feishm@cnu.edu.cn (corresponding author)}
\affiliation{School of Mathematical Sciences, Capital Normal University, Beijing 100048, China}
\affiliation{Max-Planck-Institute for Mathematics in the Sciences, Leipzig 04103, Germany}

\begin{abstract}
We study the trade-off relations satisfied by the genuine tripartite nonlocality in multipartite quantum systems. From the reduced three-qubit density matrices of the four-qubit generalized GHZ states and W states, we find that there exists a trade-off relation among the mean values of the Svetlichny operators
associated with these reduced states. Namely, the genuine three-qubit nonlocalities are not independent.
For four-qubit generalized GHZ states and W states, the summation of all their three-qubit maximal (squared) mean values of the Svetlichny operator is upper bounded. And this bound is superior to the one derived from the upper bounds of individual three-qubit mean values of the Svetlichny operator. Detailed examples are presented to illustrate the trade-off relation among the three-qubit nonlocalities.
\end{abstract}

\maketitle


\section{Introduction}
\label{sec:introd}

Nonlocality is a fundamental feature of quantum mechanics \cite{PhysRev.47.777, RevModPhys.38.447}. It is also a key resource in information processing \cite{RevModPhys.82.665,PhysRevA.81.052334,PhysRevA.61.022117, Zhao2018}, and is related to various topics in quantum information theory such as the understanding of classical and quantum boundary \cite{PhysRevLett.123.110502,ccm11}, the entangling power of nonlocal unitary operations \cite{cy14,cy16b,cy16}, and the efficient decomposition for realization in quantum circuits \cite{cy15}, unextendible product basis \cite{kai=lma}, and positive-partial-transpose entangled states \cite{cd12}.

Bell inequalities and nonlocality have been widely studied and are shown to be related to the monogamy trade-off obeyed by bipartite Bell correlations. It is believed that for general translation invariant systems, two-qubit states should not violate the Bell inequality \cite{de_Oliveira_2012}. A nontrivial model is constructed to confirm that the Bell inequality can be violated in perfect translation-invariant systems with an even number of sites \cite{Sun2013Violation}.
Monogamy relations between the violations of Bell's inequalities have been derived in \cite{PhysRevLett.102.030403}. Meanwhile, using the Bloch vectors, a trade-off relation has been derived, together with a complete classification of four-qudit quantum states \cite{Li_2019}.

In the multipartite case, nonlocality displays a much richer and more complex structure compared with the case of bipartite systems. This makes the study and the characterization of multipartite nonlocal correlations an interesting, but challenging problem. It comes thus to no surprise that our understanding of nonlocality in the multipartite setting is much less advanced than in the bipartite case \cite{PhysRevA.98.042317, PhysRevA.98.012321}.

In \cite{PhysRevLett.108.020502} a complete characterization of entanglement of an entire class of mixed three-qubit states with the same symmetry as the Greenberger-Horne-Zeilinger state, known as GHZ-symmetric states, has been achieved. By analytical expressions of maximum violation value of most efficient Bell inequalities one obtains the conditions of standard nonlocality and genuine nonlocality of this class of states. The relation between entanglement and nonlocality has been also discussed for this class of states. Interestingly, genuine entanglement of GHZ-symmetric states is necessary to reveal the standard nonlocality \cite{PhysRevA.94.032101}. Nonlocal correlations are proposed in three-qubit generalized GHZ states and four-qubit generalized GHZ states \cite{Ghose2010Multiqubit}.  Meanwhile, all multipartite pure states that are equivalent to the N-qubit W states under stochastic local operation and classical communication (SLOCC) can be uniquely determined (among arbitrary states) from their bipartite marginals \cite{PhysRevA.84.052331}.

Two overlapping bipartite binary Bell inequalities cannot be simultaneously violated, which would contradict the usual no-signaling principle. It is known as the monogamy of Bell inequality violations.
Generally Bell monogamy relations refer to trade-offs between simultaneous violations of multiple inequalities. The genuine multipartite nonlocality, as evidenced by a generalized Svetlichny inequality, does exhibit monogamy property \cite{PhysRevA.98.022133}. There is a complementarity relation between dichotomic observables leading to the monogamy of Bell inequality violations \cite{Kurzynski2010Monogamy}.

To study the nonlocality of bipartite quantum states, one considers the Clauser-Horne-Shimony-Holt (CHSH) inequality \cite{PhysRevLett.23.880}. For any two-qubit density matrix $\rho$, if there exist local hidden variable models to describe the system, the CHSH inequality says that
\begin{equation}\label{19p}
|Tr(\rho B_{CHSH})|\leq 2,
\end{equation}
where $B_{CHSH}$ is the CHSH operator
\begin{equation}\nonumber B_{CHSH}=\vec{a}\cdot\vec{\sigma}\otimes (\vec{b}+\vec{b}^{\prime}) \cdot\vec{\sigma}+\vec{a}^{\prime}\cdot \vec{\sigma}\otimes(\vec{b} -\vec{b}^{\prime })\cdot\vec{\sigma},
\end{equation}
with $\vec{a}$, $\vec{a}^{\prime}$, $\vec{b}$ and $\vec{b}^{\prime}$ the real three-dimensional unit vectors, and $\vec{\sigma}=(\sigma_{1},\sigma_{2},\sigma_{3})$ the Pauli matrices.
Denote $T$ the matrix with entries given by $t_{ij}=\tr[\rho(\sigma_{i}\otimes\sigma_{j}]$.
It has been shown that the maximal violation of the CHSH inequality (\ref{19p}) is given by \cite{HORODECKI1995340,HORODECKI1996223},
\begin{equation}\nonumber
\langle CHSH\rangle _{\rho}=\max|\mathop{\rm Tr}(\rho B_{CHSH})|=2\sqrt{M(\rho)},
\end{equation}
where $M(\rho)=max_{j<k}\{\mu_{j}+\mu_{k}\}$, $j,k\in\{1,2,3\}$, $\mu_{j},\mu_{k}$ are the two largest eigenvalues of the real symmetric matrix $T^{t}T$, and $t$ denotes the matrix transposition.

The distribution of nonlocality in multipartite systems based on the violation of Bell inequality has been investigated in \cite{PhysRevA.92.062339,PhysRevLett.118.010401}. For any 3-qubit state $\rho_{ABC}\in \mathcal{H}^A\otimes \mathcal{H}^B\otimes \mathcal{H}^C$, the maximal violation of CHSH inequality of pairwise bipartite states satisfies the following trade-off relation:
\be
\langle CHSH\rangle^2_{\rho_{AB}}+\langle CHSH\rangle^2_{\rho_{AC}}+\langle CHSH\rangle^2_{\rho_{BC}}\leq 12.
\ee
It implies that for a three qubit system, it is impossible that all pairs of qubit states violate the CHSH inequality simultaneously.

For genuine tripartite nonlocality, consider three separated observers Alice, Bob and Charlie, with their measurement settings $x, y, z$ and outputs $a, b, c$, respectively. The correlations are said to be local if the joint probability distribution $p(abc|xyz)$ can be written as
\begin{equation}
p(abc|xyz)=\int d \lambda~ q(\lambda)p_{\lambda}(a|x)p_{\lambda}(b|y)p_{\lambda}(c|z),
\end{equation}
where $\lambda$ is the local random variable and $\int d\lambda q(\lambda)=1$. A state is called genuine tripartite non-local if  $p(abc|xyz)$ can not be written as
\be\label{bi-lhv}
\bg
p(abc|xyz)=&\int d\lambda ~ q(\lambda)p_{\lambda}(ab|xy)p_{\lambda}(c|z)\\&+ \int d\mu~ q(\mu)p_{\mu}(bc|yz)p_{\mu}(a|x)\\&+\int d\nu~  q(\nu)p_{\nu}(ac|xz)p_{\nu}(b|y),
\eg
\ee
where $\int d\lambda~ q(\lambda)+\int d\mu~ q(\mu)+\int d\nu~ q(\nu)=1$. A state satisfying (\ref{bi-lhv}) is said to admit bi-LHV (local hidden variable) model. Svetlichny introduced an inequality to verify the genuine tripartite nonlocality. There are also two alternative definitions of n-way nonlocality, and a series of Bell-type inequalities for the detection of three-way nonlocality \cite{PhysRevA.88.014102}. Nevertheless, such $n-$ way nonlocalities are strictly weaker than the Svetlichny's.
The dynamics of the nonlocality measured by the violation of Svetlichny's Bell-type inequality has been investigated in the non-Markovian model \cite{Qiu_2011}.

To quantify the nonlocality of three-qubit states, in \cite{PhysRevA.96.042323}, a technique is developed to find the maximal violation of the Svetlichny inequality, and a tight upper bound is obtained. In this paper, we explicitly quantify the genuine tripartite nonlocality of the reduced states of four-qubit pure states. We first introduce the Svetlichny inequality whose violation is a signature of the genuine tripartite nonlocality. According to the maximal value of the Svetlichny operator we show that there exists a trade-off relation among the mean values of the Svetlichny operators associated with the three-qubit reduced states of GHZ and W states. We present detailed examples to illustrate the trade-off relation among such genuine three-qubit nonlocalities.
The rest of this paper is organized as follows. In Sec. \ref{sec:sye} we introduce the Svetlichny inequality. In Sec. \ref{sec:symmetric} and \ref{sec:symmetric=w} we investigate the trade-off for four-qubit symmetric pure states in the space spanned by Dicke states. Finally we conclude in Sec. \ref{sec:con}.
	
\section{Svetlichny inequality}
\label{sec:sye}

We consider the nonlocality test scenario for three-qubit systems associated with Alice, Bob and Chalie. Let the two measurement observables for Alice be $A=\vec{a}\cdot \vec{\sigma}$ and $ A^{\prime}=\vec{a}^{\prime}\cdot {\vec{\sigma}}$, where $\vec{a}$ and $\vec{a}^{\prime}$ are unit vectors in $\bbR^3$, and $\vec{\sigma}=(\sigma_{1},\sigma_{2},\sigma_{3})$ is the vector of Pauli matrices.
Each observable is an Hermitian operator with eigenvalues $\pm 1$. Similarly, we have
$B=\vec{b}\cdot \vec{\sigma}$ and $ B^{\prime}=\vec{b}^{\prime}\cdot \vec{\sigma}$ for Bob, and $C=\vec{c}\cdot \vec{\sigma}$ and $ C^{\prime}=\vec{c}^{\prime}\cdot \vec{\sigma}$ for Charlie.The Svetlichny operator corresponding to measurements $A$, $A^{\prime}$, $B$, $B^{\prime}$, $C$ and $C^{\prime}$ is defined by
\begin{equation}
\begin{aligned}\label{defS}
S:=&A\big((B+B^{\prime}) C+(B-B^{\prime}) C^{\prime}\big)\\&+A^{\prime}\big((B-B^{\prime}) C-(B+B^{\prime}) C^{\prime}\big)\\=&A(D C+D^{\prime} C^{\prime})+A^{\prime}(D^{\prime} C-D C^{\prime}),
\end{aligned}
\end{equation}
where $D=B+B^{\prime}$ and $D^{\prime}=B-B^{\prime}$.

If a 3-qubit state $\rho$ admits a bi-LHV model, then it satisfies the Svetlichny inequality \cite{PhysRevD.35.3066},
\begin{equation}
\langle S(\rho)\rangle =\mathop{\rm Tr} (S \rho)\leq 4,
\end{equation}
for all possible Svetlichny operators $S$. Conversely, a 3-qubit state which violates this inequality for some $S$ is genuine three-qubit nonlocal. To quantify the nonlocality of a 3-qubit system, we need to compute the maximum of the so-called Svetlichny value,
\begin{equation}
S_{max}(\rho)=\max\mathop{\rm Tr}(S\rho),
\end{equation}
where the maximization is taken over all possible Svetlichny operators. Thus, $S_{max}(\rho)>4$ is a sufficient condition for $\rho$ to be genuine three-qubit nonlocal. Moreover, the maximal Svetlichny value is $4\sqrt{2}$ when the Svetlichny inequality is maximally violated by, say, the GHZ state ~$(|000\rangle+|111\rangle)/\sqrt{2}$ \cite{PhysRevD.35.3066,PhysRevA.70.060101}.
It has been shown in \cite{PhysRevA.96.042323} that for any three-qubit state $\rho$, the maximal value $S_{max}$ related to the Svetlichny operator $S$ satisfies
\be\label{4lambda}
S_{\max}(\rho)\leq 4\lambda_{1},
\ee
where $\lambda_1$ is the maximum singular value of the matrix
$M=(m_{j,ik})$, with ~$m_{ijk}=\tr(\rho(\sigma_{i}\otimes\sigma_{j}\otimes\sigma_{k}))$, $i,j,k=1,2,3$.

\section{Trade-off Relations with respect to four-qubit symmetric states}
\label{sec:symmetric}

Let $\vec{x}= (\sin\theta_{x}\cos\phi_{x}, \sin\theta_{x}\sin\phi_{x}, \cos\theta_{x})$ for $x=a,a^{\prime},\,b,b^{\prime},\,c,c^{\prime}$. Set $\vec{b}+\vec{b}^{\prime} =2\vec{d} \cos\omega$ and $\vec{b}-\vec{b}^{\prime} =2\vec{d}^{\prime} \sin\omega$.
If $\omega\neq\pi n/2$ for $n\in\bbZ$, $\vec{b}+ \vec{b}^{\prime}$ and $\vec{b}-\vec{b}^{\prime}$ are mutually orthogonal.
If $\omega=\pi n/2$ for $n\in\bbZ$, for example, $\omega=\pi/2$, then $\vec{d}^{\prime}=\vec{b}$. We can still construct a $\vec{d}$ which is orthogonal to $\vec{d}^{\prime}$ in this case. These two vectors $\vec{d}$ and $\vec{d}^{\prime}$ satisfy
\be
\vec{d}\cdot \vec{d}^{\prime}= \cos\theta_{d} \cos\theta_{d^{\prime}}+\sin\theta_{d} \sin\theta_{d^{\prime}}\cos(\phi_{d}-\phi_{d^{\prime}})=0,
\ee
that is, the maximum of $\cos^{2}\theta_{d}+\cos^{2}\theta_{d^{\prime}}$ is $1$, while the maximum of $\sin^{2}\theta_{d}+\sin^{2}\theta_{d^{\prime}}$ is $2$.
Then setting $D=\vec{d}\cdot \vec{\sigma}$ and $D^{'}=\vec{d}^{'}\cdot \vec{\sigma}$, we have
\be\label{proof}
\bg
\langle S(\rho)\rangle=&2\Big|\cos\omega\langle ADC\rangle_{\rho}+\sin\omega\langle AD^{\prime}C^{\prime}\rangle_{\rho}\\&~~~~ +\sin\omega\langle A^{\prime}D^{\prime}C\rangle_{\rho}-\cos\omega\langle A^{\prime}DC^{\prime}\rangle_{\rho}\Big|\\\leq& 2\Big|\big(\langle ADC\rangle_{\rho}^{2}+\langle AD^{\prime}C^{\prime}\rangle^{2}_{\rho}\big)^{1/2}\\&~~~~+\big(\langle A^{\prime}D^{\prime}C\rangle^{2}_{\rho}+\langle A^{\prime}DC^{\prime}\rangle^{2}_{\rho}\big)^{1/2}\Big|,
\eg
\ee
where the following inequality has been taken into account,
\be\label{triangle}
x\cos\omega+y\sin\omega\leq(x^2+y^2)^{1/2},
\ee
with the equality holds when $\tan\omega=\frac{y}{x}$, \hbox{$x\cos\omega\geq 0, x\neq 0$;} or $\sin\omega=\pm 1$, \hbox{$y\sin\omega\geq 0, x=0$.} (\ref{proof}) will be used in the following derivations.

Let us consider the four-qubit generalized Greenberger-Horne-Zeilinger (GGHZ) state $\ket{\psi_{abcd}}$ and the generalized maximal slice (MS) state $|\phi_{abcd}\rangle$:
\be
\begin{split}
    |\psi_{abcd}\rangle&=\cos\theta |0000\rangle+\sin \theta |1111\rangle,\\[2mm]
    |\phi_{abcd}\rangle&=\frac{1}{\sqrt{2}} |0000\rangle+\frac{1}{\sqrt{2}} |111\rangle(\cos\theta |0\rangle+\sin \theta |1\rangle).
\end{split}
\ee
Denote $\Psi_{abcd}=|\psi_{abcd}\rangle\langle\psi_{abcd}|$ and $\Phi_{abcd}=|\phi_{abcd}\rangle\langle\phi_{abcd}|$ the corresponding density matrices.

\textbf{Theorem 1}
\label{thm:gghz}
For four-qubit GGHZ state $\Psi_{abcd}=\proj{\psi_{abcd}}$, the violation of the Svetlichny inequality on any three-qubit states satisfies the following relation:
\be\label{theorem1}
\langle S(\Psi_{abc})\rangle+\langle S(\Psi_{abd})\rangle+\langle S(\Psi_{acd})\rangle+\langle S(\Psi_{bcd})\rangle\leq16|\cos2\theta|,
\ee
where $\Psi_{abc}=\Psi_{abd}=\Psi_{acd} =\Psi_{bcd}=\cos^{2}\theta|000\rangle\langle000| +\sin^{2}\theta|111\rangle\langle111|$ are the corresponding reduced three-qubit states. The equality holds in (\ref{theorem1}) when
$$\Big|\cos\theta_{a}\cos\theta_{c}-\cos\theta_{a^{\prime}}\cos\theta_{c^{\prime}}\Big|=2,~ \omega=\theta_{d}=0,~ \theta_{d^{\prime}}=\pi/2.$$

\begin{proof}~By straightforward computation, we have
\bea
\langle ADC\rangle_{\Psi_{abc}}=\cos 2 \theta \cos \theta _{a} \cos \theta _c \cos
   \theta _d,
\eea
and similar expressions for $\langle AD^{\prime}C^{\prime}\rangle_{\Psi_{abc}}$, $\langle A^{\prime}D^{\prime}C\rangle_{\Psi_{abc}}$ and $\langle A^{\prime}DC^{\prime}\rangle_{\Psi_{abc}}$.
From (\ref{proof}), we have
\begin{eqnarray}
\begin{aligned}
\langle S(\Psi_{abc})\rangle=&2\Big|\cos\omega\langle ADC\rangle_{\Psi_{abc}}+\sin\omega\langle AD^{\prime}C^{\prime}\rangle_{\Psi_{abc}} \\&+\sin\omega\langle A^{\prime}D^{\prime}C\rangle_{\Psi_{abc}}-\cos\omega\langle A^{\prime}DC^{\prime}\rangle_{\Psi_{abc}}\Big|\\[2mm]
\leq& 2\Big|\big(\langle ADC\rangle^{2}_{\Psi_{abc}}+\langle AD^{\prime}C^{\prime}\rangle^{2}_{\Psi_{abc}}\big)^{1/2} \\&+\big(\langle A^{\prime}D^{\prime}C\rangle^{2}_{\Psi_{abc}}+\langle A^{\prime}DC^{\prime}\rangle^{2}_{\Psi_{abc}}\big)^{1/2}\Big|\\[2mm]
=&2\big|\cos2\theta\cos\theta_{a}(\cos^{2}\theta_{c} \cos^{2}\theta_{d}\\&~~~~~~~~~~~~~~~~~~~~~+\cos^{2}\theta_{c^{\prime}} \cos^{2}\theta_{d^{\prime}})^{\frac{1}{2}}\\
&+ \cos2\theta\cos\theta_{a^{\prime}}(\cos^{2}\theta_{c} \cos^{2}\theta_{d^{\prime}}\\&~~~~~~~~~~~~~~~~~~~~~+\cos^{2}\theta_{c^{\prime}} \cos^{2}\theta_{d})^{\frac{1}{2}}\big|.
\end{aligned}
\end{eqnarray}
Since the maximum of $\cos^{2}\theta_{d}+\cos^{2}\theta_{d^{\prime}}$ is 1 \cite{PhysRevLett.102.250404},
the above formula can be further reduced to be,
\be\label{th1}
\bg
\langle S(\Psi_{abc})\rangle&\leq 2\big|\cos 2\theta\big|\big(|\cos\theta_{a}|
+|\cos\theta_{a^{\prime}}|\big)\leq 4\big|\cos2\theta\big|.
\eg
\ee
Since $\langle S(\Psi_{abc})\rangle=\langle S(\Psi_{abd})\rangle=\langle S(\Psi_{acd})\rangle=\langle S(\Psi_{bcd})\rangle\leq 4\big|\cos2\theta\big|$ for the state $\Psi_{abcd}=\proj{\psi_{abcd}}$, one gets
the inequality (\ref{theorem1}).
\end{proof}

The equality holds in (\ref{theorem1}) when
$$\Big|\cos\theta_{a}\cos\theta_{c}-\cos\theta_{a^{\prime}}\cos\theta_{c^{\prime}}\Big|=2,~ \omega=\theta_{d}=0,~ \theta_{d^{\prime}}=\pi/2,$$
namely, $S_{\max}(\Psi_{abc})=S_{\max}(\Psi_{abd})=S_{\max}(\Psi_{acd})=S_{\max}(\Psi_{bcd}) =4\big|\cos2\theta\big|\leq 4$. It means that in this case all the reduced states of GGHZ state do not violate the Svetlichny inequality.

\begin{figure}[htb]
	\includegraphics[scale=0.45,angle=0]{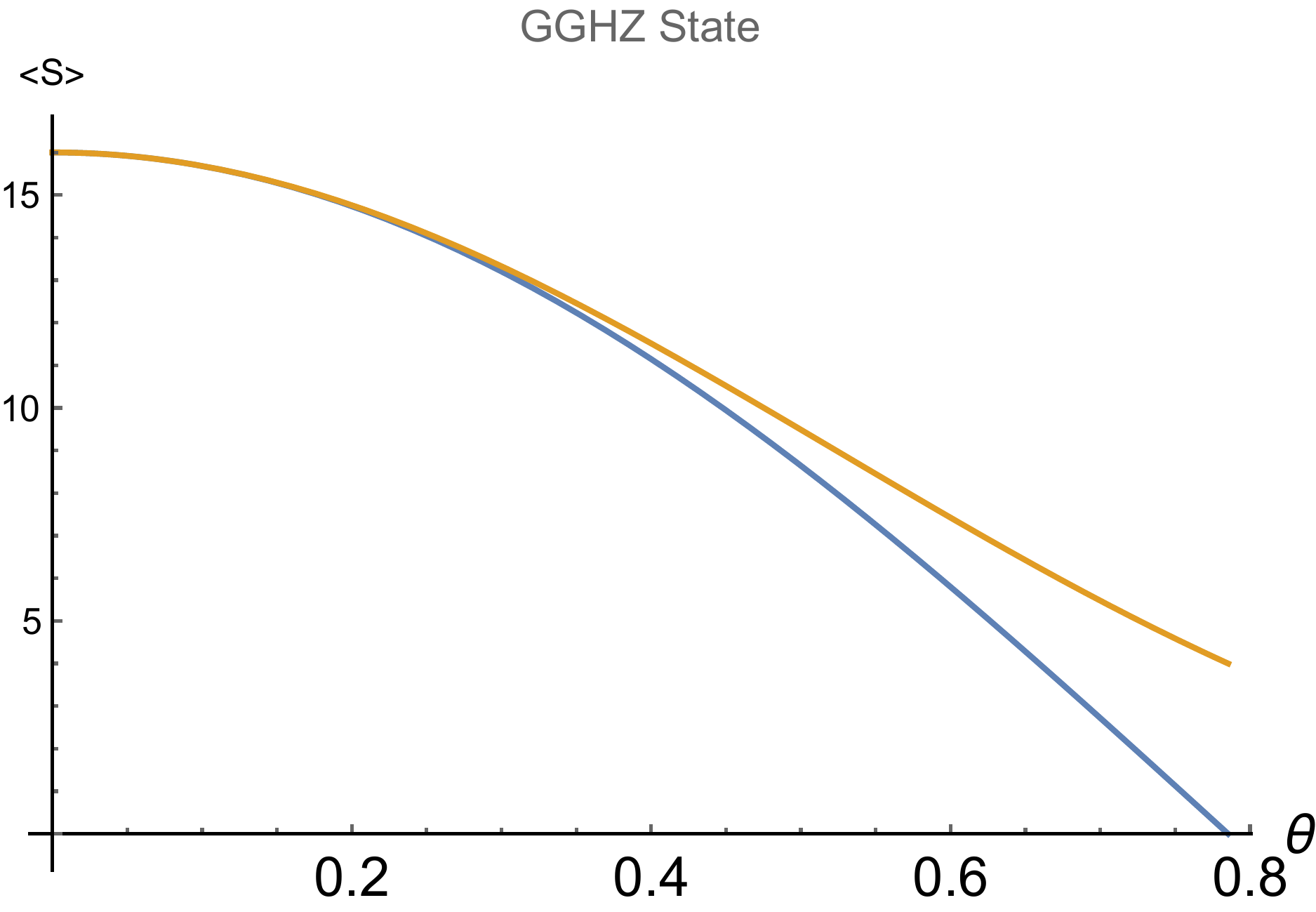}
	\caption{For $\theta\in[0,\frac{\pi}{4}]$, the upper bound of the sum of violations of the Svetlichny inequality for four reduced three-qubit states is $16|\cos2\theta|$. It is less or equal to $16\max\{\cos^{4}\theta,\sin^{4}\theta\}=16 \cos^{4}\theta$ derived from (\ref{4lambda}). The blue line is the bound from Theorem 1. The yellow one comes from (\ref{4lambda}). When $\theta=0$, two bounds are equal.}
	\label{fig:gghz}
\end{figure}

For the GGHZ state, the four reduced three-qubit states are the same.
From (\ref{4lambda}), the maximal value of the Svetlichny operator is $4\max\{\cos^{4}\theta,\sin^{4}\theta\}$ for any one of such reduced three-qubit states. 
It is remarkable that the upper bound in (\ref{theorem1}) is always less or equal to the upper bound $16\max\{\cos^{4}\theta,\sin^{4}\theta\}$ derived from (\ref{4lambda}), see Figure \ref{fig:gghz} for $\theta\in[0,\frac{\pi}{4}]$.

Generalizing Theorem 1 to general $n$-qubit case, we have for $n\geq 4$

\textbf{Corollary 1}
	For $n$-qubit GGHZ state $\ket{\Psi}=\cos\theta\ket{00\cdots0}+\sin\theta\ket{11\cdots1}$, the violation of the Svetlichny inequality on any three-qubit states satisfies the following relation:
	\be
	\sum_{1\leq I<J<K\leq n} \langle S(\Psi_{IJK})\rangle \leq 4 {n\choose 3}|\cos2\theta|,
	\ee
where $\Psi_{IJK}={\tr}_{\overline{IJK}}\proj{\Psi}=\cos^{2}\theta|000\rangle\langle000|_{IJK} +\sin^{2}\theta|111\rangle\langle111|_{IJK}$ are the corresponding reduced three-qubit states associated with qubits $I$, $J$ and $K$, and ${\tr}_{\overline{IJK}}$ stands for the trace over the rest qubit systems.

\textbf{Theorem 2}\label{thm:ms}
For four-qubit generallized MS states $\Phi_{abcd}$,
the violation of the Svetlichny inequality on the reduced three-qubit density matrices satisfies the following relation:
\be\label{theorem2}
\bg
&\langle S(\Phi_{abc})\rangle+\langle S(\Phi_{abd})\rangle+\langle S(\Phi_{acd})\rangle+\langle S(\Phi_{bcd})\rangle\\&\leq 4\sqrt{2}|\cos\theta|+12|\cos^2\theta+\frac{1}{2}\sin2\theta|,
\eg
\ee
where
\be
\bg
\Phi_{abc}=&\frac{1}{2}|000\rangle\langle000| +\frac{1}{2}\cos\theta|000\rangle\langle111|\\&+ \frac{1}{2}\cos\theta|111\rangle\langle000|+ \frac{1}{2}|111\rangle\langle111|, \\
\Phi_{abd}=&\Phi_{acd}=\Phi_{bcd}\\ =&\frac{1}{2}|000\rangle\langle000| +\frac{1}{2}\cos^{2}\theta|110\rangle\langle110|\\&+\frac{1}{2}\cos\theta \sin\theta|110\rangle\langle111|+\frac{1}{2}\cos\theta \sin\theta|111\rangle\langle110|\\&+\frac{1}{2}\sin^{2}\theta|111\rangle\langle111|.
\eg
\ee

See	proof in Appendix A.

\begin{figure}[htb]
	\includegraphics[scale=0.45,angle=0]{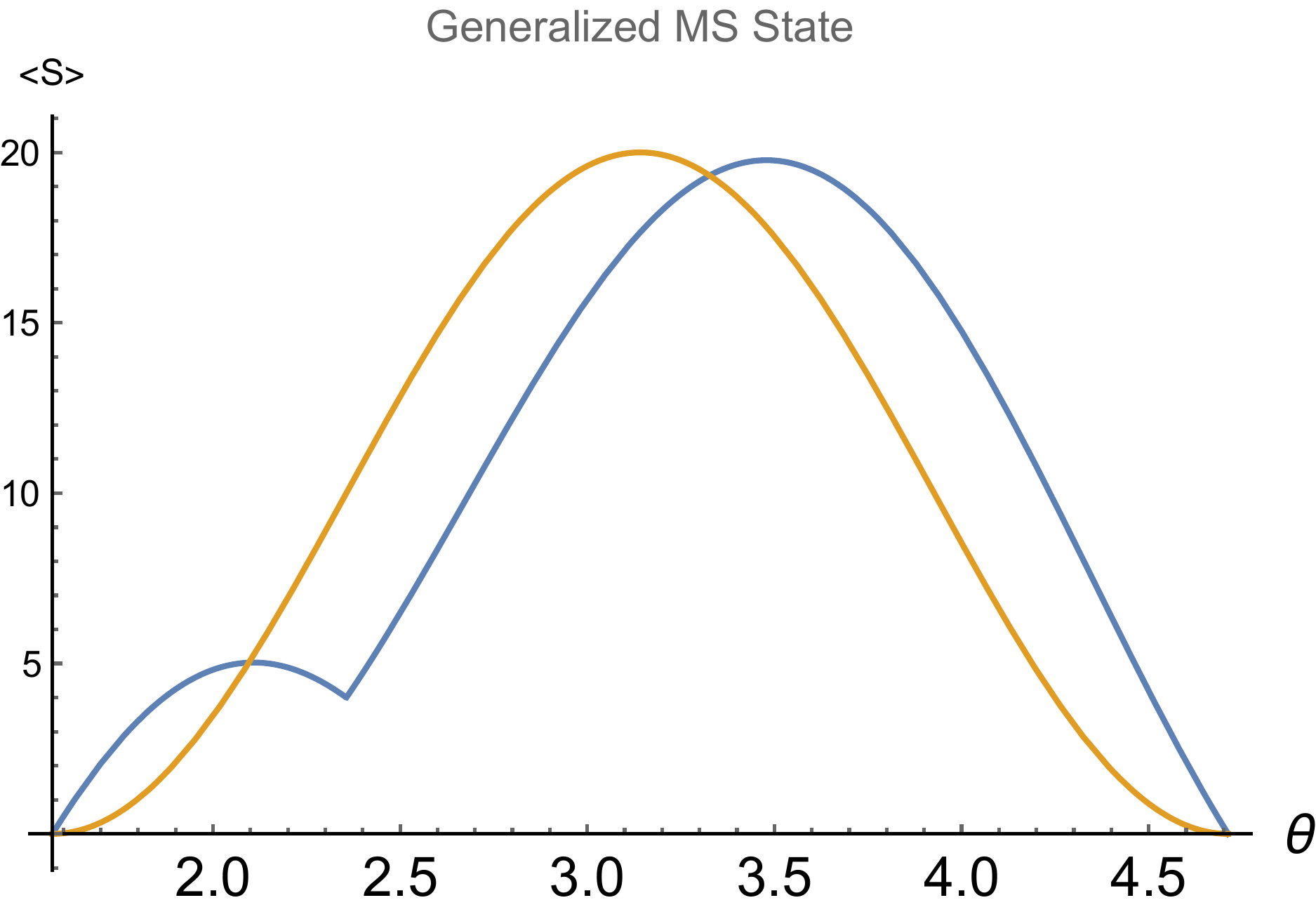}
	\caption{The blue line is for the upper bound of (\ref{theorem2}), the yellow line is for the upper bound of (\ref{c4}) for $\theta\in (\pi/2, 3\pi/2)$.}
	\label{fig:MS}
\end{figure}

Inequality (\ref{theorem2}) gives a trade off relation of among the three-qubit genuine nonlocalities
in MS states. In fact, by using (\ref{4lambda}) for any three-qubit states of a MS state, one has
\be\label{c4}
\langle S(\Phi_{abc})\rangle+\langle S(\Phi_{abd}) \rangle+\langle S(\Phi_{acd})\rangle+\langle S(\Phi_{bcd})\rangle\leq 20\cos^2{\theta}.
\ee
Nevertheless, the upper bound of (\ref{c4}) is larger than the one of (\ref{theorem2}),
see Figure \ref{fig:MS} for $\theta \in (\pi/2, 3\pi/2)$.

Now consider the $n$-qubit generalized MS states,
\bea
|\Psi_{12\dots n}\rangle&=\frac{1}{\sqrt{2}} |00\cdots0\rangle+\frac{1}{\sqrt{2}} |11\cdots1\rangle\ket{\psi},
\eea
where $\ket{\psi}=\cos\theta |0\rangle+\sin \theta |1\rangle$.
Let $\mathcal{I}$ denote a proper subset of $\{1,2,\dots,n\} $.
We define the states with $n\notin\mathcal{I}$ the Class I ($\#\mathcal{I}=m<n$), and $n\in\mathcal{I}$ the Class II. Then, there are $({n\choose m}-{n\choose m-1})$ states in class I, and ${n\choose m-1}$ states in class II,
	\be
	\rho_{\mathcal{I}}=\begin{cases}
		\frac{1}{2}\proj{0\cdots00}+\frac{1}{2}\proj{1\cdots11} & n\notin\mathcal{I}\\
		\frac{1}{2}\proj{0\cdots00}+\frac{1}{2}\proj{1\cdots1\psi} & n\in\mathcal{I}
	\end{cases}
	\ee

From Theorem 2 we have the following corollary,

\textbf{Corollary 2}\label{nms1}
	For $n-$qubit generallized MS states $\Phi_{abcd}$,
	the violation of the Svetlichny inequality on the reduced three-qubit density matrices satisfies the following relation:
	\be
	\bg
	&\sum_{1\leq I<J<K\leq n} \langle S(\Psi_{IJK})\rangle \\& \leq 4\sqrt{2}{n-1\choose 2}|\cos\theta|\\&+4\Big({n\choose 3}-{n-1\choose 2}\Big)|\cos^2\theta+\frac{1}{2}\sin2\theta|,
	\eg
	\ee
where $\Psi_{IJK}={\tr}_{\overline{IJK}}\proj{\Psi}=\frac{1}{2}|000\rangle\langle000|_{IJK} +\frac{1}{2}|111\rangle\langle111|_{IJK}$ for $\Psi_{IJK}$ belonging to Class I, and $\Psi_{IJK}={\tr}_{\overline{IJK}}\proj{\Psi}=\frac{1}{2}|000\rangle\langle000| +\frac{1}{2}\cos^{2}\theta|110\rangle\langle110|+\frac{1}{2}\cos\theta \sin\theta|110\rangle\langle111|+\frac{1}{2}\cos\theta \sin\theta|111\rangle\langle110|+\frac{1}{2}\sin^{2}\theta|111\rangle\langle111|$ for $\Psi_{IJK}$ belonging to Class II.

\section{Trade-off Relations for the W-class States}
\label{sec:symmetric=w}

For a 4-qubit state:
\begin{equation}\label{abcd}
\begin{aligned}
\ket{\varphi}_{abcd}&=\alpha|1000\rangle+\beta|0100\rangle+\gamma |0010\rangle+\delta|0001\rangle+\lambda\ket{0000},
\end{aligned}
\end{equation}
with $\alpha, \b, \g, \d, \l$ are real numbers. It can generate four-qubit quantum states by unitary operators. We consider trade-off relation between the reduced states of $\ket{\psi_{abcd}}$.

\textbf{Theorem 3}\label{th:w}
For any 4-qubit state $\ket{\varphi}_{abcd}$, the violation of Svetlichny operators on tripartite states satisfies the following relation:
	\be\label{eqn3}
	\bg
	&\langle S(\rho_{abc})\rangle+\langle S(\rho_{abd})\rangle+\langle S(\rho_{acd})\rangle+\langle S(\rho_{bcd})\rangle
	\\[2mm]&\leq 2\Big((2x_1+8y_1)^{\frac{1}{2}}+(2x_1+8y_1+8\b^2\g^2)^{\frac{1}{2}}\Big)\\&+2\Big((2x_2+8y_2)^{\frac{1}{2}}+(2x_2+8y_2+8\b^2\d^2)^{\frac{1}{2}}\Big)\\[2mm]&+2\Big((2x_3+8y_3)^{\frac{1}{2}}+(2x_3+8y_3+8\d^2\l^2)^{\frac{1}{2}}\Big)\\[2mm]&+2\Big((2x_4+8y_4)^{\frac{1}{2}}+(2x_4+8y_4+8\d^2\g^2)^{\frac{1}{2}}\Big).
	\eg
	\ee
	where $\rho_{abc}={\tr_{d}}\proj{\varphi}_{abcd}$, $\rho_{abd}={\tr_{c}}\proj{\varphi}_{abcd}$, $\rho_{acd}={\tr_{b}}\proj{\varphi}_{abcd}$, $\rho_{bcd}={\tr_{a}}\proj{\varphi}_{abcd}$, and
	\begin{equation*}
	\begin{aligned}
	x_1&=(\a^2+\b^2+\g^2-\d^2-\l^2)^2,\\ y_1&=\b^2\g^2+\a^2\l^2+\frac{3}{2}\a^2\b^2+\g^2\l^2+\frac{3}{2}\a^2\g^2+\b^2\l^2,\\
	x_2&=(\a^2+\b^2-\g^2+\d^2-\l^2)^2,\\ y_2&=\b^2\g^2+\a^2\l^2+\frac{3}{2}\a^2\b^2+\d^2\l^2+\frac{3}{2}\a^2\d^2+\d^2\b^2,\\
	x_3&=(\a^2-\b^2+\g^2+\d^2-\l^2)^2,\\ y_3&=\frac{3}{2}\a^2\d^2+\a^2\l^2+\frac{3}{2}\a^2\g^2+\d^2\l^2+\d^2\g^2+\l^2\g^2,\\
	x_4&=(-\a^2+\b^2+\g^2+\d^2-\l^2)^2,\\ y_4&=\frac{3}{2}\b^2\g^2+\b^2\l^2+\d^2\g^2+\d^2\l^2+\g^2\l^2+\frac{3}{2}\d^2\b\g.
	\end{aligned}
	\end{equation*}

See proof in Appendix B.

The $n$-qubit Dicke state is an $n$-partite symmetric state defined as $\ket{\mathfrak{D}(n,m)}={n\choose m}^{-1/2}\sum_{P\in\cP} P(\ket{0}^{\otimes m}\ox \ket{1}^{\otimes (n-m)})$, where $\cP$ is the permutation group of $n$ elements. The state $\ket{\mathfrak{D}(4,1)}$ is the standard 4-qubit W state.
When $\lambda=0$, the state (\ref{abcd}) reduces to the 4-qubit W-class state:
\begin{equation}
\begin{aligned}
|\varphi\rangle_{W_{abcd}}=\alpha|1000\rangle+\beta|0100\rangle +\gamma|0010\rangle +\delta|0001\rangle.
\end{aligned}
\end{equation}
For the state (\ref{eqn3}) reduces to
	\be\label{eqn3p}
	\bg
	&\langle S({W_{abc}})\rangle^{2}+\langle S({W_{abd}})\rangle^{2}+\langle S({W_{acd}})\rangle^{2}+\langle S({W_{bcd}})\rangle^{2}\\[2mm]
	&\leq 64(1+\alpha^2\gamma^2 +\beta^2\delta^2+2\alpha^2\beta^2+2\beta^2\gamma^2 +2\gamma^2\delta^2),
	\eg
	\ee
where $W_{abc}$, $W_{abd}$, $W_{acd}$ and $W_{bcd}$ denote the corresponding reduced states of
$|\varphi\rangle \langle\varphi|_{W_{abcd}}$.

However, from (\ref{4lambda})
the violation of Svetlichny operators for tripartite states $W_{abc}$, $W_{abd}$, $W_{acd}$ and $W_{bcd}$
satisfy the following relations,
\be
\begin{split}
\langle S({W_{abc}})\rangle
&\leq 4\max\{\sqrt{4(\alpha\beta^2+\alpha\gamma^{2})}, \sqrt{8\beta\gamma^2+(2\delta^{2}-1) ^{2}} \},\\[2mm]
\langle S({W_{abd}})\rangle
&\leq 4\max\{\sqrt{4(\alpha\beta^2+\alpha\delta^{2})}, \sqrt{8\beta\delta^2+(2\gamma^{2}-1) ^{2}} \},\\[2mm]
\langle S({W_{acd}})\rangle
&\leq 4\max\{\sqrt{4(\alpha\gamma^2+\alpha\delta^{2})}, \sqrt{8\gamma\delta^2+(2\beta^{2}-1) ^{2}} \},\\[2mm]
\langle S({W_{bcd}})\rangle
&\leq 4\max\{\sqrt{4(\beta\gamma^2+\beta\delta^{2})}, \sqrt{8\gamma\delta^2+(2\alpha^{2}-1) ^{2}} \}.
\end{split}
\ee
Accounting to the fact that $\max\{X,Y\}=\frac{\mid X-Y\mid+\mid X+Y\mid}{2}$, one has
\be\label{wstatere}
\bg
&\langle S({W_{abc}})\rangle^{2}+\langle S({W_{abd}})\rangle^{2}+\langle S({W_{acd}})\rangle^{2}+\langle S({W_{bcd}})\rangle^{2}\\\leq&8(\mid 4(\alpha\beta^2+\alpha\gamma^{2}) -8\beta\gamma^2-(2\delta^{2}-1) ^{2}\mid\\ &+\mid 4(\alpha\beta^2+ \alpha\delta^{2})-8\beta\delta^2-(2\gamma^{2}-1) ^{2}\mid\\&+\mid 4(\beta\gamma^2+\beta\delta^{2})- 8\gamma\delta^2-(2\alpha^{2}-1) ^{2}\mid\\&+\mid 4(\alpha\gamma^2+\alpha\delta^{2})-8\gamma\delta^2-(2\beta^{2}-1) ^{2}\mid\\&+ 8(\alpha\beta^2+\alpha\gamma^{2}+ \alpha\delta^{2}+\frac{3}{2}\beta\gamma^2+\frac{3}{2}\beta\delta^2+ 2\gamma\delta^2)\\&+(2\alpha^{2}-1) ^{2}+(2\beta^{2}-1) ^{2}+(2\gamma^{2}-1) ^{2}+(2\delta^{2}-1) ^{2}).
\eg
\ee

\begin{figure*}[htbp!]
	\centering
	\begin{tabular}{@{}cccc@{}}
		\includegraphics[width=0.45\textwidth]{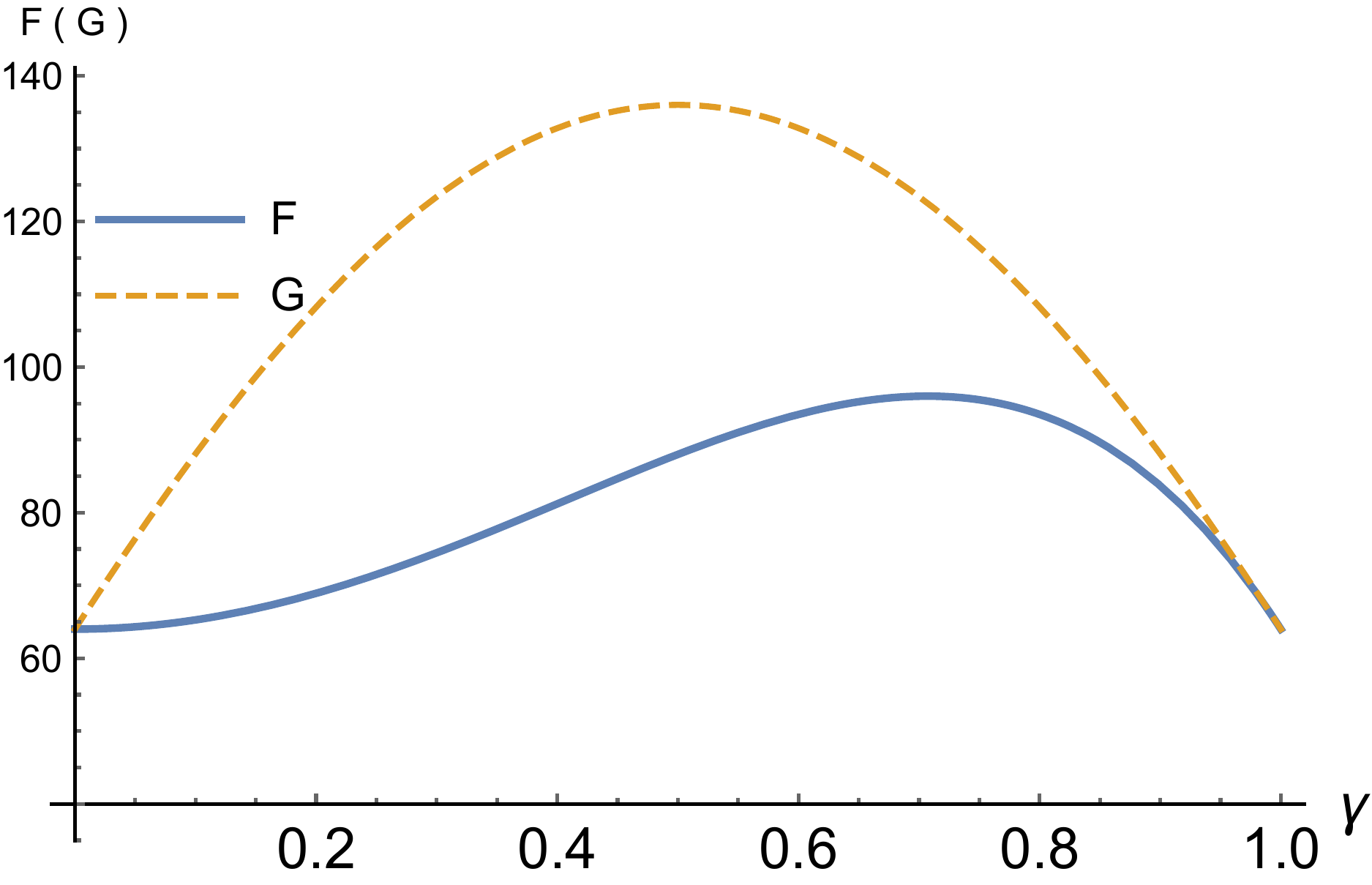}
	\end{tabular}
	\caption{(Color online) In the range of $\gamma\in [0,1]$, the bound $F$ is smaller than $G$.}
	\label{x=0plot1}
\end{figure*}

Denote $F$ and $G$ the right sides of (\ref{eqn3p}) and (\ref{wstatere}), respectively.
Figure \ref{x=0plot1} shows that
the value of $F$ is always less than $G$ in the range $\gamma\in [0,1]$ for
$\alpha=\beta=0$ and $\delta^2=1-\gamma^2$.

\begin{figure*}[t]
	{\centering
		\begin{tabular}{ccc}
			\hspace{-0.3 cm}
			\resizebox*{0.5\textwidth}{0.26\textheight}{\includegraphics{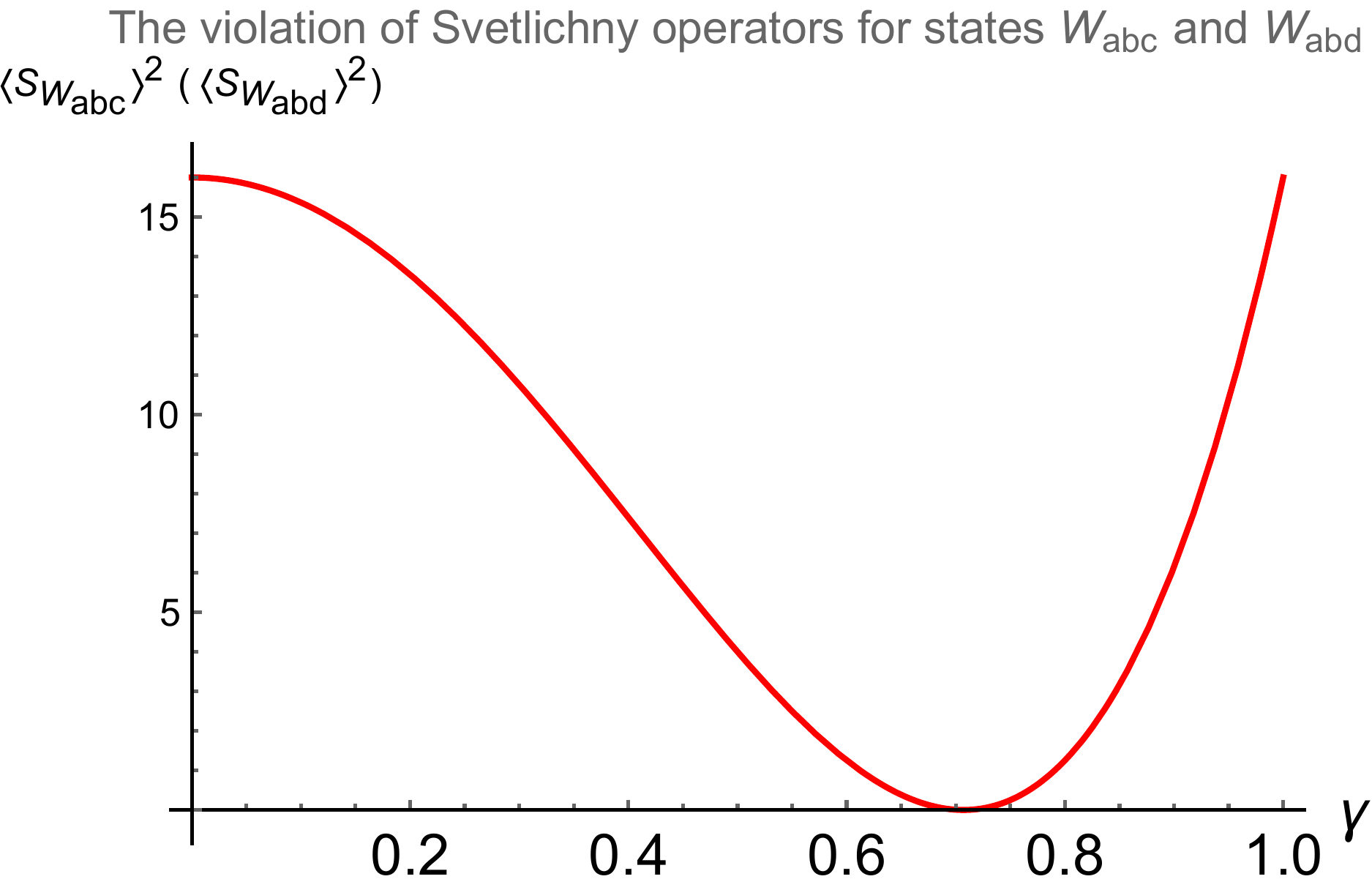}}
			& & \hspace{-0.3 cm}
			\resizebox*{0.5\textwidth}{0.26\textheight}{\includegraphics{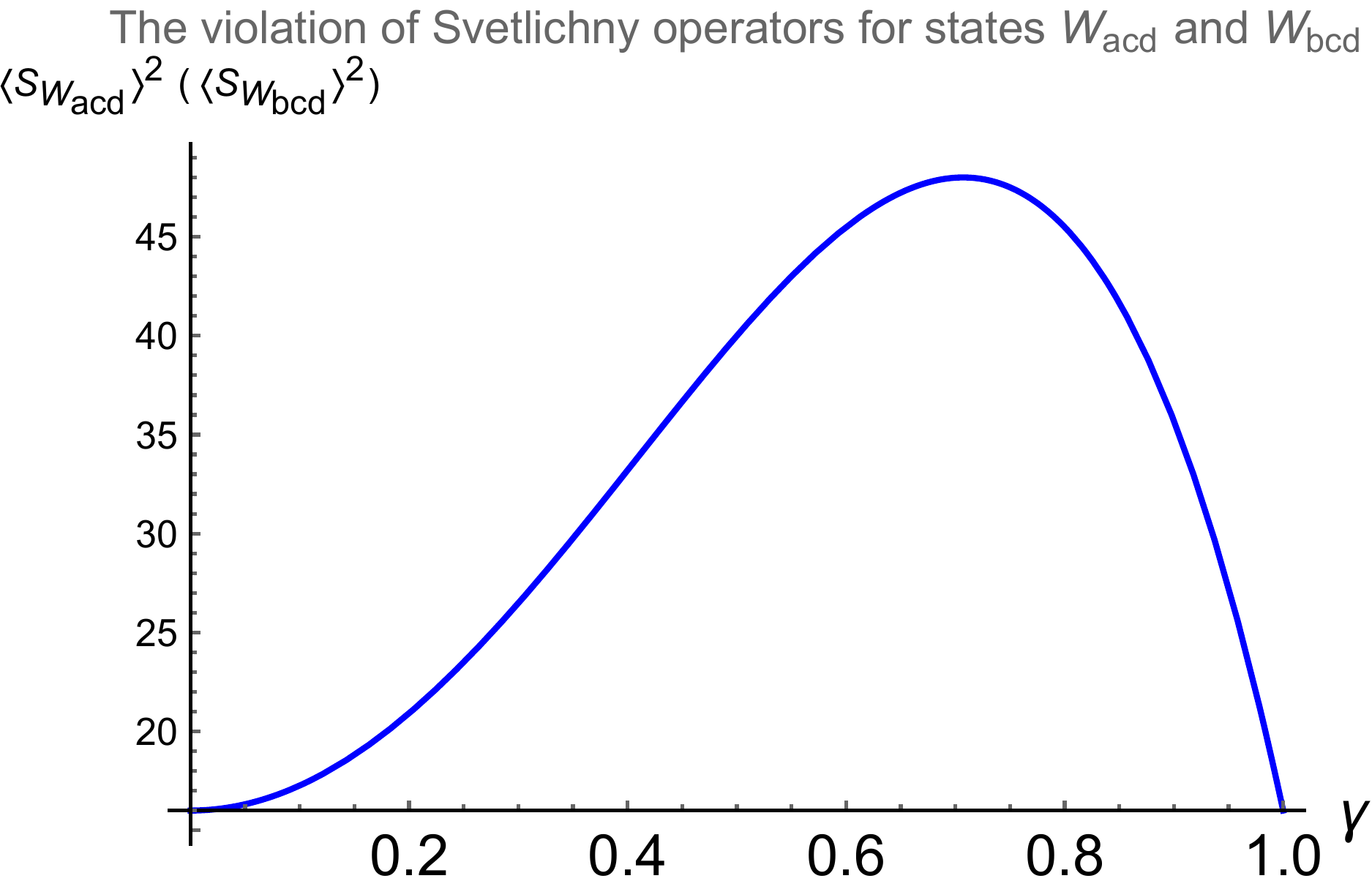}} \vspace{2ex}\\
		\end{tabular}\par
	\centering
	\begin{tabular}{@{}cccc@{}}
		\includegraphics[width=0.50\textwidth]{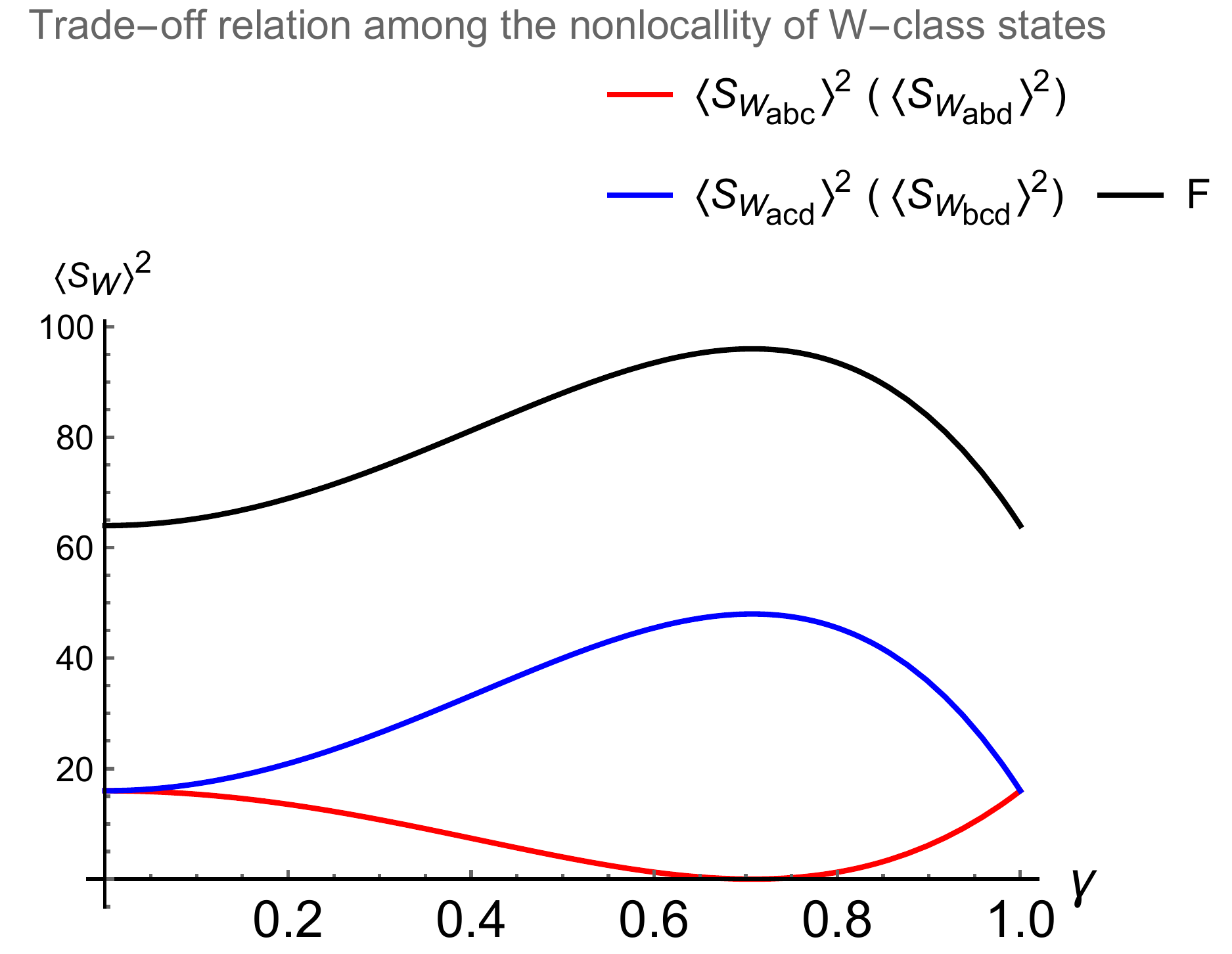}
	\end{tabular}
	\caption{(Color online) In $\gamma\in [0,1]$, the quantities
		$\langle S_{{W_{abc}}}\rangle^2(=\langle S_{{W_{abd}}}\rangle^2)$ and
		$\langle S_{{W_{acd}}}\rangle^2(=\langle S_{{W_{bcd}}}\rangle^2)$ vary in a way such that
		their summation kept to be bounded by $F$.}
	\label{x=0tradeoff}}
\end{figure*}

Equation (\ref{eqn3p}) also gives a kind of trade-off relation among the quantum nonlocality of the reduced states. The maximum value $704/7$ of $F$ is attained at  $\{\alpha, \beta, \gamma, \delta\}=\{0, \sqrt{2/7} \sqrt{3/7}, \sqrt{2/7}\}$. Figure \ref{x=0tradeoff} shows the detailed trade off relations
among $\langle S_{{W_{abc}}}\rangle^2$,  $\langle S_{{W_{abd}}}\rangle^2$, $\langle S_{{W_{acd}}}\rangle^2$ and $\langle S_{{W_{bcd}}}\rangle^2$.
Here for $\alpha=\beta=0$ and $\delta^2=1-\gamma^2$, we have
$\langle S_{{W_{abc}}}\rangle^2=\langle S_{{W_{abd}}}\rangle^2$ and
$\langle S_{{W_{acd}}}\rangle^2=\langle S_{{W_{bcd}}}\rangle^2$.

\section{Conclusions}
\label{sec:con}

We have studied the trade-off relationship of genuine tripartite non-locality in a multipartite system.
And the corresponding tight upper bounds for GHZ-class states and W-class states have presented, showing that the genuine three-qubit nonlocalities are not independent in a four-qubit system.
Meanwhile, we have identified that the reduced three-qubit states of a four-qubit GHZ state can not violate the Svetchlity inequality. Our approach may be also used to investigate the trade off relations of genuine nonlocalities satisfied by the reduced tripartite states of a more general multipartite system.

\bigskip
\medskip

\section*{Acknowledgements}

The work was initiated when the first author was a Phd student in Capital Normal University. LJZ, LC, KW, and YS are supported by the NNSF of China (Grant No. 11871089) and the Fundamental Research Funds for the Central Universities (Grant Nos. KG12080401 and ZG216S1902).
SMF and YMG acknowledges the support from NSF of China under Grant No. 11675113 and Key Project of Beijing Municipal Commission of Education (KZ201810028042).

\bibliography{studyoncorrelation}

\begin{thebibliography}{10}

\bibitem{PhysRev.47.777}
A.~Einstein, B.~Podolsky, and N.~Rosen.
\newblock Can quantum-mechanical description of physical reality be considered
  complete?
\newblock {\em Phys. Rev.}, 47:777--780, May 1935.

\bibitem{RevModPhys.38.447}
JOHN~S. BELL.
\newblock On the problem of hidden variables in quantum mechanics.
\newblock {\em Rev. Mod. Phys.}, 38:447--452, Jul 1966.

\bibitem{RevModPhys.82.665}
Harry Buhrman, Richard Cleve, Serge Massar, and Ronald de~Wolf.
\newblock Nonlocality and communication complexity.
\newblock {\em Rev. Mod. Phys.}, 82:665--698, Mar 2010.

\bibitem{PhysRevA.81.052334}
Ashok Ajoy and Pranaw Rungta.
\newblock Svetlichny's inequality and genuine tripartite nonlocality in
  three-qubit pure states.
\newblock {\em Phys. Rev. A}, 81:052334, May 2010.

\bibitem{PhysRevA.61.022117}
Asher Peres.
\newblock Classical interventions in quantum systems. ii. relativistic
  invariance.
\newblock {\em Phys. Rev. A}, 61:022117, Jan 2000.

\bibitem{Zhao2018}
Li-Jun Zhao, Yu-Min Guo, XianQing Li-Jost, and Shao-Ming Fei.
\newblock Quantum nonlocality can be distributed via separable states.
\newblock {\em Science China Physics, Mechanics {\&} Astronomy}, 61(7):70321,
  Mar 2018.

\bibitem{PhysRevLett.123.110502}
Fernando G. S.~L. Brandao, Elizabeth Crosson, M.~Burak~Sahinoglu Crosson, and
  John Bowen.
\newblock Quantum error correcting codes in eigenstates of
  translation-invariant spin chains.
\newblock {\em Phys. Rev. Lett.}, 123:110502, Sep 2019.

\bibitem{ccm11}
Lin Chen, Eric Chitambar, Kavan Modi, and Giovanni Vacanti.
\newblock Detecting multipartite classical states and their resemblances.
\newblock {\em Phys. Rev. A}, 83:020101, Feb 2011.

\bibitem{cy14}
Lin Chen and Li~Yu.
\newblock {Nonlocal and controlled unitary operators of Schmidt rank three}.
\newblock {\em Phys. Rev. A}, 89:062326, Jun 2014.

\bibitem{cy16b}
Lin Chen and Li~Yu.
\newblock Entangling and assisted entangling power of bipartite unitary
  operations.
\newblock {\em Phys. Rev. A}, 94:022307, Aug 2016.

\bibitem{cy16}
Lin Chen and Li~Yu.
\newblock Entanglement cost and entangling power of bipartite unitary and
  permutation operators.
\newblock {\em Phys. Rev. A}, 93:042331, Apr 2016.

\bibitem{cy15}
Lin Chen and Li~Yu.
\newblock Decomposition of bipartite and multipartite unitary gates into the
  product of controlled unitary gates.
\newblock {\em Phys. Rev. A}, 91:032308, Mar 2015.

\bibitem{kai=lma}
Kai Wang, Lin Chen, Yi~Shen, Yize Sun, and Li-Jun Zhao.
\newblock Constructing $2 \times 2 \times 4$ and $4 \times 4$ unextendible
  product bases and positive-partial-transpose entangled states.
\newblock {\em Linear and Multilinear Algebra}, 0(0):1--16, 2019.

\bibitem{cd12}
Lin Chen and Dragomir~Z. Djokovic.
\newblock Qubit-qudit states with positive partial transpose.
\newblock {\em Phys. Rev. A}, 86:062332, Dec 2012.

\bibitem{de_Oliveira_2012}
Thiago~R. de~Oliveira, A.~Saguia, and M.~S. Sarandy.
\newblock Nonviolation of bell‘s inequality in translation invariant systems.
\newblock {\em {EPL} (Europhysics Letters)}, 100(6):60004, dec 2012.

\bibitem{Sun2013Violation}
Zhao~Yu Sun, Yu~Ying Wu, Hai~Lin Huang, Bo~Jun Chen, and Bo~Wang.
\newblock Violation of bell inequality in perfect translation invariant
  systems.
\newblock {\em Physical Review A}, 88(88):340--347, 2013.

\bibitem{PhysRevLett.102.030403}
Marcin Pawlowski and \ifmmode \check{C}\else~\v{C}\fi{}aslav Brukner.
\newblock Monogamy of bell's inequality violations in nonsignaling theories.
\newblock {\em Phys. Rev. Lett.}, 102:030403, Jan 2009.

\bibitem{Li_2019}
Ming Li, Zong Wang, Jing Wang, Shuqian Shen, and Shao ming Fei.
\newblock The norms of bloch vectors and classification of four-qudits quantum
  states.
\newblock {\em {EPL} (Europhysics Letters)}, 125(2):20006, feb 2019.

\bibitem{PhysRevA.98.042317}
Ming-Xing Luo.
\newblock Nonlocality of all quantum networks.
\newblock {\em Phys. Rev. A}, 98:042317, Oct 2018.

\bibitem{PhysRevA.98.012321}
R.~Augusiak, M.~Demianowicz, and J.~Tura.
\newblock Constructing genuinely entangled multipartite states with
  applications to local hidden variables and local hidden states models.
\newblock {\em Phys. Rev. A}, 98:012321, Jul 2018.

\bibitem{PhysRevLett.108.020502}
Christopher Eltschka and Jens Siewert.
\newblock Entanglement of three-qubit greenberger-horne-zeilinger--symmetric
  states.
\newblock {\em Phys. Rev. Lett.}, 108:020502, Jan 2012.

\bibitem{PhysRevA.94.032101}
Biswajit Paul, Kaushiki Mukherjee, and Debasis Sarkar.
\newblock Nonlocality of three-qubit greenberger-horne-zeilinger--symmetric
  states.
\newblock {\em Phys. Rev. A}, 94:032101, Sep 2016.

\bibitem{Ghose2010Multiqubit}
S.~Ghose, S.~Debnath, N.~Sinclair, A.~Kabra, and R.~Stock.
\newblock Multiqubit nonlocality in families of 3- and 4-qubit entangled
  states.
\newblock {\em Journal of Physics A Mathematical \& Theoretical},
  43(44):445301, 2010.

\bibitem{PhysRevA.84.052331}
Swapan Rana and Preeti Parashar.
\newblock Optimal reducibility of all $w$ states equivalent under stochastic
  local operations and classical communication.
\newblock {\em Phys. Rev. A}, 84:052331, Nov 2011.

\bibitem{PhysRevA.98.022133}
Ravishankar Ramanathan and Piotr Mironowicz.
\newblock Trade-offs in multiparty bell-inequality violations in qubit
  networks.
\newblock {\em Phys. Rev. A}, 98:022133, Aug 2018.

\bibitem{Kurzynski2010Monogamy}
P.~Kurzynski, T.~Paterek, R.~Ramanathan, W.~Laskowski, and D.~Kaszlikowski.
\newblock Monogamy of multipartite bell inequality violations.
\newblock {\em Physical Review Letters}, 106(106):180402, 2010.

\bibitem{PhysRevLett.23.880}
John~F. Clauser, Michael~A. Horne, Abner Shimony, and Richard~A. Holt.
\newblock Proposed experiment to test local hidden-variable theories.
\newblock {\em Phys. Rev. Lett.}, 23:880--884, 1969.

\bibitem{HORODECKI1995340}
R.~Horodecki, P.~Horodecki, and M.~Horodecki.
\newblock Violating $\mathrm{B}$ell inequality by mixed spin-1$/$2 states:
  necessary and sufficient condition.
\newblock {\em Phys. Lett. A}, 200(5):340--344, 1995.

\bibitem{HORODECKI1996223}
R.~Horodecki.
\newblock Two-spin-1$/$2 mixtures and $\mathrm{B}$ell's inequalities.
\newblock {\em Phys. Lett. A}, 210(4):223--226, 1996.

\bibitem{PhysRevA.92.062339}
HuiHui Qin, ShaoMing Fei, and Xianqing Li-Jost.
\newblock Trade-off relations of $\mathrm{B}$ell violations among pairwise
  qubit systems.
\newblock {\em Phys. Rev. A}, 92:062339, 2015.

\bibitem{PhysRevLett.118.010401}
Shuming Cheng and Michael J.~W. Hall.
\newblock Anisotropic invariance and the distribution of quantum correlations.
\newblock {\em Phys. Rev. Lett.}, 118:010401, 2017.

\bibitem{PhysRevA.88.014102}
Jean-Daniel Bancal, Jonathan Barrett, Nicolas Gisin, and Stefano Pironio.
\newblock Definitions of multipartite nonlocality.
\newblock {\em Phys. Rev. A}, 88:014102, Jul 2013.

\bibitem{Qiu_2011}
Liang Qiu.
\newblock Nonlocality sudden birth and transfer in system and environment.
\newblock {\em Chinese Physics Letters}, 28(3):030301, mar 2011.

\bibitem{PhysRevA.96.042323}
Ming Li, Shuqian Shen, Naihuan Jing, ShaoMing Fei, and Xianqing Li-Jost.
\newblock Tight upper bound for the maximal quantum value of the
  $\mathrm{S}$vetlichny operators.
\newblock {\em Phys. Rev. A}, 96:042323, 2017.

\bibitem{PhysRevD.35.3066}
George Svetlichny.
\newblock Distinguishing three-body from two-body nonseparability by a
  $\mathrm{B}$ell-type inequality.
\newblock {\em Phys. Rev. D}, 35:3066--3069, 1987.

\bibitem{PhysRevA.70.060101}
Peter Mitchell, Sandu Popescu, and David Roberts.
\newblock Conditions for the confirmation of three-particle nonlocality.
\newblock {\em Phys. Rev. A}, 70:060101, 2004.

\bibitem{PhysRevLett.102.250404}
S.~Ghose, N.~Sinclair, S.~Debnath, P.~Rungta, and R.~Stock.
\newblock Tripartite entanglement versus tripartite nonlocality in three-qubit
  $\mathrm{G}$reenberger-$\mathrm{H}$orne-$\mathrm{Z}$eilinger-$\mathrm{C}$lass
  states.
\newblock {\em Phys. Rev. Lett.}, 102:250404, 2009.

\end{thebibliography}
\bibliographystyle{unsrt}

\appendix
\section{ Proof of Theorem 2}

\begin{widetext}

For the reduced state $\Phi_{abc}$, one has the expectation value of the Svetlichny operator,
		$$
		\langle ADC\rangle_{\Phi_{abc}}=\cos \theta  \sin \theta_{a} \sin \theta
		_{c} \sin\theta _{d} \cos \left(\phi _a+\phi _c+\phi
		_d\right).
		$$
		$\langle AD^{\prime}C^{\prime}\rangle_{\Phi_{abc}}$, $\langle A^{\prime}D^{\prime}C\rangle_{\Phi_{abc}}$ and
		$\langle A^{\prime}DC^{\prime}\rangle_{\Phi_{abc}}$ have similar expressions.
			Therefore we have
			
			\be\small\label{th2}
			\bg
			\langle S(\Phi_{abc})\rangle=&2\Big|\cos\omega\langle ADC\rangle_{\Phi_{abc}}+\sin\omega\langle AD^{\prime}C^{\prime}\rangle_{\Phi_{abc}} +\sin\omega\langle A^{\prime}D^{\prime}C\rangle_{\Phi_{abc}}-\cos\omega\langle A^{\prime}DC^{\prime}\rangle_{\Phi_{abc}}\Big|\\[2mm]\leq&2\Big|\big(\langle ADC\rangle^{2}_{\Phi_{abc}}+\langle AD^{\prime}C^{\prime}\rangle^{2}_{\Phi_{abc}}\big)^{1/2} +\big(\langle A^{\prime}D^{\prime}C\rangle^{2}_{\Phi_{abc}}+\langle A^{\prime}DC^{\prime}\rangle^{2}_{\Phi_{abc}}\big)^{1/2}\Big|\\[2mm]
			\leq&2|\{\big(\cos\theta\sin \theta_{a} \sin \theta
			_{c} \sin\theta _{d} \cos \left(\phi _a+\phi _c+\phi
			_d\right)\big)^2+\big(\cos\theta\sin \theta _{a} \sin\theta
			_{c^{\prime}} \sin\theta
			_{d^{\prime}} \cos \left(\phi
			_{c^{\prime}}+\phi
			_{d^{\prime}}+\phi _{a}\right)\big)^2\}^{1/2}\\[2mm]&+\{\big(\cos\theta\sin \theta _{c} \sin \theta
			_{a^{\prime}} \sin \theta
			_{d^{\prime}} \cos \left(\phi
			_{a^{\prime}}+\phi
			_{d^{\prime}}+\phi _{c}\right)\big)^2+\big(\cos\theta\cos\omega\sin \theta _{d} \sin \theta
			_{a^{\prime}} \sin \theta
			_{c^{\prime}} \cos \left(\phi
			_{a^{\prime}}+\phi
			_{c^{\prime}}+\phi _{d}\right)\big)^2\}^{1/2}|\\[2mm]
			\leq&2|{(\cos^2\theta\sin^2\theta_{d} +\cos^2\theta\sin^2\theta_{d^{\prime}})}^{1/2} +{(\cos^2\theta\sin^2\theta_{d^{\prime}} +\cos^2\theta\sin^2\theta_{d})}^{1/2}|\\[2mm]
			\leq&4|\cos\theta(\sin^2\theta_{d}+\sin^2\theta_{d^{\prime}}) ^{1/2}|\\[2mm]
			\leq&4\sqrt{2}|\cos\theta|.
			\eg
			\ee
			
			When $\phi_{i}+\phi_{j}+\phi_{k}=0$, where $i\in\{a,a^{\prime}\}$, $j\in\{d,d^{\prime}\}$ and $k\in\{c,c^{\prime}\}$, one has $\langle S(\Phi_{abc})\rangle=4\sqrt{2}|\cos\theta|$.
			
			For the reduced state $\Phi_{abd}$, we have
			\bea
			\langle ADC\rangle_{\Phi_{abd}}=\frac{1}{2} \cos \theta _a \cos \theta _d \left(\sin 2\theta
			\sin \theta _c \cos\phi _c+2 \cos
			^2\theta \cos \theta _c\right).
			\eea
			The expressions for  $\langle AD^{\prime}C^{\prime}\rangle_{\Phi_{abd}}$, $\langle A^{\prime}D^{\prime}C\rangle_{\Phi_{abd}}$ and $\langle A^{\prime}DC^{\prime}\rangle_{\Phi_{abd}}$ are similar.
			By direct computation we obtain
			
			\bea
			\begin{aligned}
				\langle S(\Phi_{abd})\rangle=&2\Big|\cos\omega\langle ADC\rangle_{\Phi_{abd}}+\sin\omega\langle AD^{\prime}C^{\prime}\rangle_{\Phi_{abd}} +\sin\omega\langle A^{\prime}D^{\prime}C\rangle_{\Phi_{abd}}-\cos\omega\langle A^{\prime}DC^{\prime}\rangle_{\Phi_{abd}}\Big|\\[2mm]
				\leq&2\Big|\big(\langle ADC\rangle^{2}_{\Phi_{abd}}+\langle AD^{\prime}C^{\prime}\rangle^{2}_{\Phi_{abd}}\big)^{1/2} +\big(\langle A^{\prime}D^{\prime}C\rangle^{2}_{\Phi_{abd}}+\langle A^{\prime}DC^{\prime}\rangle^{2}_{\Phi_{abd}}\big)^{1/2}\Big|\\[2mm]
				=&2\Big|\{(\frac{1}{2} \cos \theta _a \cos \theta _d \left(\sin 2\theta
				\sin \theta _c \cos\phi _c+2 \cos
				^2\theta \cos \theta _c\right))^2\\&+(\frac{1}{2} \cos \theta _a \cos\theta_{d^{\prime}} \left(\sin 2
				\theta  \sin \theta _{c^{\prime}} \cos \phi
				_{c^{\prime}}+2 \cos ^2\theta  \cos \theta _{c^{\prime}}\right))^2\}^{1/2}\\&+\{(\frac{1}{2} \cos \theta _{a^{\prime}} \cos \theta _{d^{\prime}}
				\left(\sin 2 \theta  \sin \theta _c \cos \phi _c+2
				\cos ^2\theta  \cos \theta _c\right))^2\\&+(\frac{1}{2} \cos \theta _d \cos \theta _{a^{\prime}} \left(\sin 2\theta  \sin \theta_{c^{\prime}} \cos \phi
				_{c^{\prime}}+2 \cos ^2\theta  \cos \theta _{c^{\prime}}\right))^2\}^{1/2}\Big|\\[2mm]
				 \leq&4\big|(\frac{1}{4}\sin^{2}2\theta+\sin2\theta\cos^{2}\theta+\cos^{4}\theta)^{1/2}\big|\\[2mm]
				\leq&4|\cos^2\theta+\frac{1}{2}\sin2\theta|.
			\end{aligned}
			\eea
		
		Taking into account that $\langle S(\Phi_{abd})\rangle=\langle S(\Phi_{acd})\rangle=\langle S(\Phi_{bcd})\rangle$, one proves the Theorem.

\section{Proof of Theorem 3  }

For the reduced state $\rho_{abc}$,  \begin{equation}
		\begin{aligned}
		{\rho_{abc}}=&{\tr}_{d}\proj{\varphi}_{abcd}\\=&\alpha^{2}|100\rangle\langle 100|+\alpha\beta|100\rangle\langle010|+\alpha\gamma|100\rangle \langle001|\\&+\alpha\lambda\ket{100}\bra{000}+\alpha\beta|010\rangle\langle 100|+\beta^{2}|010\rangle\langle010|\\&+\beta\gamma|010\rangle\langle001|+\beta\lambda\ket{010}\bra{000}+\alpha\gamma|001\rangle\langle 100|\\&+\beta\gamma|001\rangle\langle010|+\gamma^{2}|001\rangle \langle001|+\gamma\lambda\ket{001}\bra{000}\\&+\delta^{2}|000\rangle\langle000|+\alpha\lambda\ket{000}\ket{100}+\beta\lambda\ket{000}\bra{010}\\&+\gamma\lambda\ket{000}\bra{001}+\lambda^2\proj{000}.
		\end{aligned}
		\end{equation}
		we can obtain
		\be
		\bg
		&\langle ADC\rangle_{{\rho _{abc}}}\\[2mm]
		=&-(\a^2+\b^2+\g^2-\s^2-\l^2)\cos \theta_a\cos \theta_c\cos \theta_d\\[2mm]
		&+2\b\g\cos(\phi_{c}-\phi_{d})\cos \theta_a\sin \theta_c\sin \theta_d \\[2mm]&+ 2\a\l\cos\phi_a \sin \theta_a\cos \theta_c\cos \theta_d  \\[2mm]
		&+2\a\b\cos(\phi_{a}-\phi_{d})\sin \theta_a\cos \theta_c\sin \theta_d \\[2mm]&+ 2\g\l\cos\phi_c \cos \theta_a\sin \theta_c\cos \theta_d   \\[2mm]
		&+2\a\g\cos(\phi_{a}-\phi_{c})\sin \theta_a\sin \theta_c\cos \theta_d \\[2mm]&+ 2\b\l\cos\phi_d \cos\theta_a\cos \theta_c\sin \theta_d.
		\eg
		\ee
		Let
		\beq
		\label{eq:coeff-1}
		\bal
		u_1&= -(\a^2+\b^2+\g^2-\s^2-\l^2), \quad v_1=0,  \\[2mm]
		u_2&= 2\b\g\cos(\phi_{c}-\phi_{d}), \quad v_2=2\a\l\cos\phi_a,  \\[2mm]
		u_3&= 2\a\b\cos(\phi_{a}-\phi_{d}), \quad v_3=2\g\l\cos\phi_c,   \\[2mm]
		u_4&= 2\a\g\cos(\phi_{a}-\phi_{c}), \quad v_4=2\b\l\cos\phi_d,
		\eal
		\eeq
		and
		\beq
		\label{eq:coeff-2}
		\bal
		x_1&=\cos \theta_a\cos \theta_c\cos \theta_d,\quad y_1&=\sin \theta_a\sin \theta_c\sin \theta_d,  \\[2mm]
		x_2&=\cos \theta_a\sin \theta_c\sin \theta_d,\quad y_2&=\sin \theta_a\cos \theta_c\cos \theta_d,   \\[2mm]
		x_3&=\sin \theta_a\cos \theta_c\sin \theta_d,\quad y_3&=\cos \theta_a\sin \theta_c\cos \theta_d,   \\[2mm]
		x_4&=\sin \theta_a\sin \theta_c\cos \theta_d,\quad y_4&=\cos \theta_a\cos \theta_c\sin \theta_d.
		\eal
		\eeq
		One can verify that $\sum_{j=1}^4(x_j^2+y_j^2)=1$. Hence we consider the following optimization:
		\beq
		\label{eq:opti}
		\max (\sum_i u_ix_i+\sum_j v_jy_j) ~~~s.t. \quad \sum_{j=1}^4(x_j^2+y_j^2)=1.
		\eeq
		Using Lagrange multiplier, we have the maximum $k=\sqrt{\sum_i (u_i^2+v_i^2)}$. It follows that the maximum is attained when each $\cos\phi=\pm 1$.

			Therefore, we have
			$$\langle ADC\rangle_{\rho_{{abc}}}\leq \sqrt{(\a^2+\b^2+\g^2-\s^2-\l^2)^2+4(\b^2\g^2+\a^2\l^2+\a^2\b^2+\g^2\l^2+\a^2\g^2+\b^2\l^2)}.$$
			
			Similarly, we have	
			\be
			\bg
			\langle AD^{\prime}C^{\prime}\rangle_{{\rho_{abc}}}
			&\leq \sqrt{(\a^2+\b^2+\g^2-\d^2-\l^2)^2+4(\b^2\g^2+\a^2\l^2+2\a^2\b^2+\g^2\l^2+2\a^2\g^2+\b^2\l^2)},\\
			\langle A^{\prime}D^{\prime}C\rangle_{{\rho _{abc}}}
			&\leq \sqrt{(\a^2+\b^2+\g^2-\d^2-\l^2)^2+4(2\b^2\g^2+\a^2\l^2+\a^2\b^2+\g^2\l^2+2\a^2\g^2+\b^2\l^2)},\\
			\langle A^{\prime}DC^{\prime}\rangle_{{\rho_{abc}}}
			&\leq \sqrt{(\a^2+\b^2+\g^2-\d^2-\l^2)^2+4(2\b^2\g^2+\a^2\l^2+2\a^2\b^2+\g^2\l^2+\a^2\g^2+\b^2\l^2)}.
			\eg
			\ee
		
		Therefore, concerning the violation of the Svetlichny inequality with respect to the reduced state ${\rho_{abc}}$
		we have
		\be
		\bg
		\langle S(\rho_{abc})\rangle=&2\big|\cos\theta\langle ADC\rangle_{\rho_{abc}}+\sin\theta\langle AD^{\prime}C^{\prime}\rangle_{\rho_{abc}}\\& +\sin\theta\langle A^{\prime}D^{\prime}C\rangle_{\rho_{abc}}-\cos\theta\langle A^{\prime}DC^{\prime}\rangle_{\rho_{abc}}\big|\\
		\leq&2\Big|\big(\langle ADC\rangle^{2}_{\rho_{{abc}}}+\langle AD^{\prime}C^{\prime}\rangle^{2}_{\rho_{{abc}}}\big)^{1/2} \\&+\big(\langle A^{\prime}D^{\prime}C\rangle^{2}_{\rho_{{abc}}}+\langle A^{\prime}DC^{\prime}\rangle^{2}_{\rho_{{abc}}}\big)^{1/2}\Big|\\[2mm]
		=&2\Big((2x_1+8y_1)^{\frac{1}{2}}+(2x_1+8y_1+8\b^2\g^2)^{\frac{1}{2}}\Big),
		\eg
		\ee
		where $x_1=(\a^2+\b^2+\g^2-\d^2-\l^2)^2$, $y_1=\b^2\g^2+\a^2\l^2+\frac{3}{2}\a^2\b^2+\g^2\l^2+\frac{3}{2}\a^2\g^2+\b^2\l^2$.
		Similarly, with respect to the reduced states $\rho_{abd}$, $\rho_{acd}$ and $\rho_{bcd}$, we get
		\begin{eqnarray}
		\begin{aligned}
		\langle S(\rho_{abd})\rangle	&\leq 2\Big((2x_2+8y_2)^{\frac{1}{2}}+(2x_2+8y_2+8\b^2\d^2)^{\frac{1}{2}}\Big),
		\end{aligned}
		\end{eqnarray}
		where $x_2=(\a^2+\b^2-\g^2+\d^2-\l^2)^2$, $y_2=\b^2\g^2+\a^2\l^2+\frac{3}{2}\a^2\b^2+\d^2\l^2+\frac{3}{2}\a^2\d^2+\d^2\b^2$.
		\begin{eqnarray}
		\begin{aligned}
		\langle S(\rho_{acd})\rangle	&\leq 2\Big((2x_3+8y_3)^{\frac{1}{2}}+(2x_3+8y_3+8\d^2\l^2)^{\frac{1}{2}}\Big),
		\end{aligned}
		\end{eqnarray}
		where $x_3=(\a^2-\b^2+\g^2+\d^2-\l^2)^2$, $y_3=\frac{3}{2}\a^2\d^2+\a^2\l^2+\frac{3}{2}\a^2\g^2+\d^2\l^2+\d^2\g^2+\l^2\g^2$.
		\begin{eqnarray}
		\begin{aligned}
		\langle S(\rho_{bcd})\rangle&\leq 2\Big((2x_4+8y_4)^{\frac{1}{2}}+(2x_4+8y_4+8\d^2\g^2)^{\frac{1}{2}}\Big),
		\end{aligned}
		\end{eqnarray}
		where $x_4=(-\a^2+\b^2+\g^2+\d^2-\l^2)^2$, $y_4=\frac{3}{2}\b^2\g^2+\b^2\l^2+\d^2\g^2+\d^2\l^2+\g^2\l^2+\frac{3}{2}\d^2\b\g$.

\end{widetext}

\end{document}